\documentclass[amsmath,amssymb,aps,prx,reprint]{revtex4-2}
\usepackage{graphicx} 
\usepackage[caption=false]{subfig}
\usepackage{placeins}
\usepackage{tikz}
\usetikzlibrary{positioning}

\usepackage{mathtools}
\DeclarePairedDelimiter\bra{\langle}{\rvert}
\DeclarePairedDelimiter\ket{\lvert}{\rangle}
\DeclarePairedDelimiterX\braket[2]{\langle}{\rangle}{#1\,\delimsize\vert\,\mathopen{}#2}
\DeclarePairedDelimiter\expval{\langle}{\rangle}

\DeclareMathOperator{\Tr}{Tr}
\DeclareMathOperator{\proj}{\mathbb{P}}

\begin{document}
\title{Partial Wavefunction Collapse Under Repeated Weak Measurement of a Non-Conserved Observable}

\author{Carter J. Swift}
\author{Nandini Trivedi}%
\affiliation{Department of Physics, The Ohio State University, Columbus, OH 43210
}%

    \begin{abstract}
    Two hallmarks of quantum non-demolition (QND) measurement are the ensemble-level conservation of the expectation value of the measured observable $A$ and the eventual, inevitable collapse of the system into some eigenstate of $A$. This requires that $A$ commutes with $H$, the system's Hamiltonian. In what we term ``Auxiliary Observable QND'' measurement, $A$ does not commute with $H$ and the above two characteristics clearly cannot be present as the system's dynamics prevent $\langle A \rangle$ from reaching a definite value. However, in this paper we find that under such a measurement QND behavior still arises, but is seen in the behavior of a secondary ``target'' observable we call $B$, with the condition that $B$ commutes with both $A$ and $H$. In such cases, the expectation value of $B$ is conserved and the system at least partially collapses with respect to eigenstates of $B$. We show as an example how this surprising result applies to a Heisenberg chain, where we demonstrate that local measurements on a single site can reveal information about the spectrum of an entire system, a finding which may be of practical use in experiments.
\end{abstract}

\maketitle

\section{Introduction}

\begin{figure*}
    \subfloat{
        \begin{tikzpicture}[node distance=0cm]
            \node (img) {\includegraphics[width=0.4\textwidth]{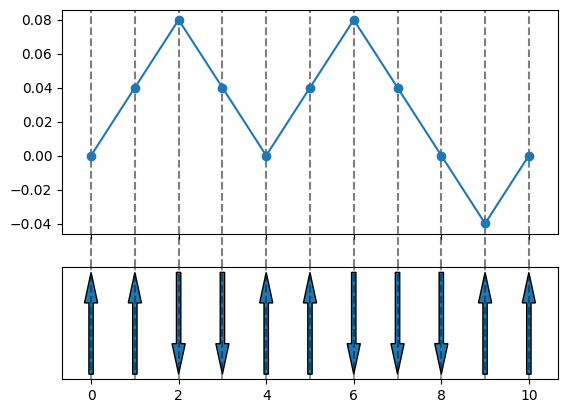}};
            \node[below=of img,yshift=0.1cm] {\footnotesize Timestep};
            \node[left=of img,xshift=0.4cm,yshift=-1.3cm,rotate=90,anchor=center] {\footnotesize Meas. Result};
            \node[left=of img,xshift=0.0cm,yshift=0.8cm,rotate=90,anchor=center] {\footnotesize $\expval{\sigma^z}$};
            \node[above left=of img]{(a)};
        \end{tikzpicture}
    }%
    \subfloat{
        \begin{tikzpicture}[node distance=0cm]
            \node (img) {\includegraphics[width=0.4\textwidth]{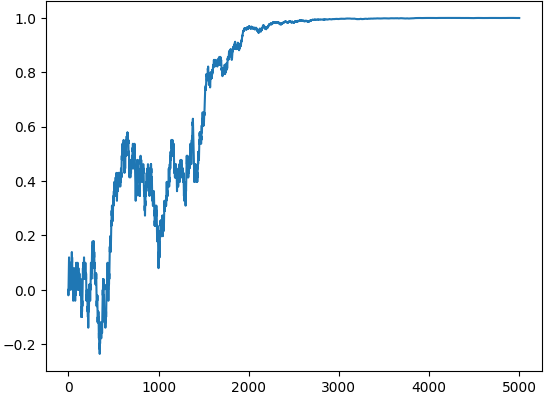}};
            \node[below=of img,yshift=0.1cm,xshift=0.3cm] {\footnotesize Timestep};
            \node[left=of img,xshift=0cm,yshift=0.2cm,rotate=90,anchor=center] {\footnotesize $\expval{\sigma^z}$};
            \node[above left=of img]{(b)};
        \end{tikzpicture}
    }
    
    \subfloat{
        \begin{tikzpicture}[node distance=0cm]
            \node (img) {\includegraphics[width=0.4\textwidth]{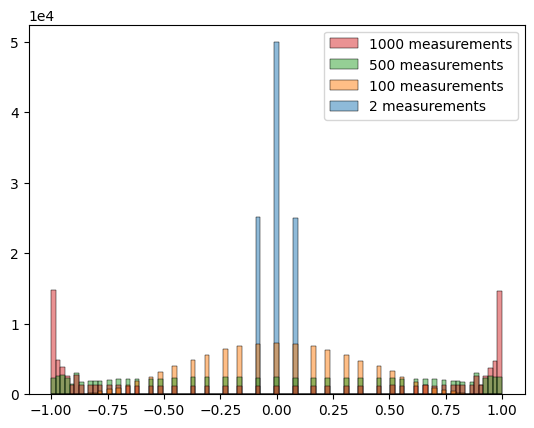}};
            \node[below=of img,yshift=0.1cm,xshift=0.1cm] {\footnotesize $\expval{\sigma^z}$};
            \node[left=of img,xshift=-0.1cm,yshift=0cm,rotate=90,anchor=center] {\footnotesize Count};
            \node[above left=of img,yshift=-0.5cm]{(c)};
        \end{tikzpicture}
    }%
    \subfloat{
    \begin{tikzpicture}[node distance=0cm]
        \node (img) {\includegraphics[width=0.4\textwidth]{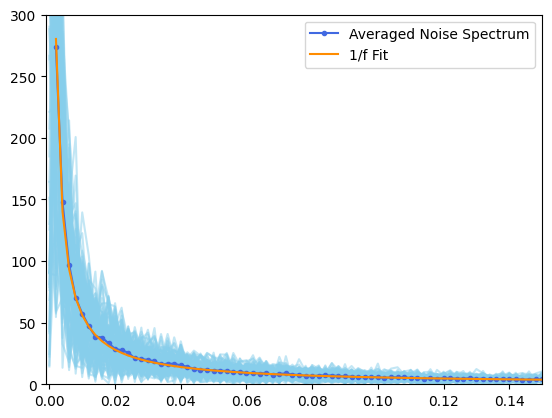}};
        \node[above left = of img, yshift=-0.4cm]{(d)};
        \node[below = of img, yshift=0.1cm,xshift=0cm]{\footnotesize Frequency};
        \node[left= of img, xshift=0.0cm, yshift=0.0cm, rotate=90, anchor=center] {\footnotesize Amplitude};
    \end{tikzpicture}
     
    }

         \caption{QND Weak Measurement of $\sigma^z$ of a qubit initialized in $\ket{\rightarrow}$. a) (Bottom) A series of weak measurement results from projection of the ancilla; (Top) The expectation value $\expval{\sigma^z}$ of the qubit corresponding to the measurement record. Each result nudges the expectation value at the following timestep in the same direction as the result. b) Repeated weak measurement of the qubit leads to eventual collapse to an eigenstate of $\sigma^z$ following some period of stochastic evolution. c) Distributions of $\expval{\sigma^z}$ for $10^5$ simulated qubits subjected to repeated weak measurement, shown after certain numbers of measurements have been completed. It is seen that while each individual run collapses to an eigenstate, the ensemble-average expectation value of 0 is conserved. (d) Averaged pre-collapse noise spectrum of 100 runs shows 1/f behavior.}
     \label{fig:qnd}
\end{figure*}


An important concept in the theory of quantum measurements is the notion of wavefunction collapse, where following a projective measurement of a system prepared in a superposition of eigenstates, any subsequent measurements of the same observable on the same system yield the same eigenvalue as the first measurement. Weak measurement is a type of measurement in which the system of interest is not directly projected in order to avoid this immediate collapse. This can be achieved by entangling the system with an ancillary degree of freedom which is itself then projectively measured. In this way, controlling the parameters coupling the ancilla to the system results in a tunable trade-off between the amount of information obtained and the degree to which the original system state is disturbed (for an overview of weak measurement, see \cite{Svensson13}). When the observable to be measured commutes with the system Hamiltonian this produces a quantum non-demolition (QND) measurement \cite{Thorne78}. Importantly, non-demolition does not mean that the system is unchanged, just that the ensemble average of the expectation value of the measured observable is unchanged by the measurement. It has been shown that repeated QND weak measurements converge to an eigenstate of the measured observable, asymptotically reproducing a projective measurement and the associated wavefunction collapse \cite{Bauer2011}. In Fig. \ref{fig:qnd}, we illustrate this behavior in the example of QND weak measurement of a single qubit. In particular, panel (c) highlights the two important properties of eventual collapse and conservation of ensemble-average expectation value. These two properties have made QND weak measurement an appealing tool for experiment, as it behaves in a way that one expects measurements to behave, while giving the experimenter more control than can be achieved with only projective measurements.

QND measurement has received significant attention over the years, particularly in the context of experiment. Early work on foundations and applications of QND measurements can be seen in \cite{Thorne78} and \cite{Braginsky80}. More recently, QND weak measurement was used in a breakthrough experiment that was able to intercept and reverse individual quantum jumps as they happened \cite{Minev19}. QND has also featured heavily in the context of understanding wavefunction collapse and the ``measurement problem'' \cite{Zurek81,Zurek82}. Collapse under weak measurement has also been examined by Korotkov in the context of continuous measurement of a double quantum dot \cite{Korotkov99}.

In contrast to QND measurement, one might instead consider a case in which the measured operator does not commute with the system Hamiltonian. For reasons that will soon become clear, we call this situation Auxiliary Observable QND (AO-QND) measurement. In ordinary QND the measurement back-action accumulates uninterrupted by any system dynamics, culminating in eventual collapse to an eigenstate of the measured operator, yet in AO-QND sequential measurement the system dynamics do interfere with the back-action, clearly rendering it impossible for collapse to an eigenstate of the measured operator to occur. So, for the behavior of a system subjected to this type of measurement, we would perhaps expect something resembling the early-time behavior of the QND case, but with no eventual collapse. One would therefore expect that subjecting a system to such a measurement protocol would not reveal any useful information. Indeed, such a protocol has been determined to behave as if the measurement basis was chosen randomly for each measurement, making gleaning useful information a challenging task, although it has also been shown that the protocol could at least be used for quantum state tomography \cite{Ashhab09}. 

In this paper we examine the consequences of AO-QND measurement processes and present as an example a numerical simulation of such a process being applied to a particular system, the Heisenberg chain. Our results indicate that the naïve expectation of non-collapse does not fully reflect reality, as an AO-QND measurement process does indeed extract information from the system and lead to at least a partial collapse of the system state. In section II we approach the problem theoretically and prove that such collapse can be expected under general circumstances for any operator which commutes with both the system Hamiltonian and the measured operator. Then, in section III we treat the particular case of a Heisenberg chain, finding that repeatedly measuring the z-component of the spin on the end of the chain causes it to oscillate with sharp Fourier peaks that correspond to energy gaps in the spectrum of the chain. We show that these oscillations arise as a result of the chain collapsing into a sector with a definite total z-component, making contact with our general proof. Finally, we comment on how these findings indicate that AO-QND measurement protocols may be of practical use in experiment and should not be overlooked.

\section{Repeated Weak Measurement}

We now provide a theoretical description of the weak measurement process, for general systems weakly measured by a sequence of qubit ancillas. We will derive all results without assuming QND, then show how the known QND results are recovered by substituting in that assumption. 

Given a system with Hamiltonian $H_0$, we want to measure some observable $A$. To do so, we first couple it to an ancilla initialized in the state $\frac{1}{\sqrt{2}}(\ket{\uparrow} + \ket{\downarrow})$ via a pre-measurement interaction $\gamma A \otimes \sigma^z_a \equiv \gamma H_m$. Readout of the ancilla is then performed by projective measurement of $\sigma^y_a$. These weak measurements are performed back-to-back in sequence, concurrently with the evolution of the system under $H_0$.


Defining $\tau$ to be the time over which the pre-measurement is performed, the joint system-ancilla density matrix $\alpha$ becomes
\begin{equation}
\alpha(\tau) = U \alpha U^\dagger \label{eq:evolve}
\end{equation}
with 
\begin{equation}
U = e^{-i(H_0 + \gamma H_{m})\tau} \label{eq:U}.
\end{equation}
The regime in which this pre-measurement yields a meaningful weak measurement is the limit $\tau << 1$ and $\gamma \tau << 1$. Thus, we expand (\ref{eq:U}) to second order in $\tau$ to obtain
\begin{equation}
    \begin{aligned}
    U = &1 - iH_0\tau - i\gamma\tau H_{m}\\ &-
    \frac{1}{2}\tau^2({H_0^2} +\gamma{H_0H_{m}} + \gamma{H_{m}H_0}+ \gamma^2{H_{m}^2}).
    \end{aligned}
    \label{eq:Utaylor}
\end{equation}
Substituting into (\ref{eq:evolve}) and keeping terms up to second order then gives
\begin{widetext}
\begin{equation}
    \begin{aligned}
        \alpha(\tau) &= \alpha + i\tau([\alpha,H_0] + \gamma[\alpha,H_{m}])
        + \tau^2 (H_0\alpha H_0 - \frac{1}{2}\{H_0^2,\alpha\} + \gamma^2H_{m}\alpha H_m
        - \frac{\gamma^2}{2}\{H_m^2,\alpha\} + \gamma H_0 \alpha H_m + \gamma H_m \alpha H_0\\
        &- \frac{\gamma}{2}\{H_0H_m,\alpha\} - \frac{\gamma}{2}\{H_mH_0,\alpha\}).
    \end{aligned}
\end{equation}
\end{widetext}
Following the pre-measurement, a projective readout of $\sigma^y_a$ is performed, yielding the result $\pm 1$ with probabilities given by
\begin{align}
    P(\pm 1) &= \frac{1}{2}(1 + \expval{\sigma^y_a}) = \frac{1}{2} + \frac{1}{2}\Tr(\sigma^y_a\alpha(\tau))\nonumber \\
    &= \frac{1}{2} \pm \gamma\tau \Tr(A\rho) \pm \frac{i\gamma\tau^2}{2}\Tr([H_0,A]\rho) \label{eq:weakprobs}
\end{align}
where $\rho$ is the density matrix of the system prior to the pre-measurement. We thus see that to leading order this probability behaves as expected for a weak measurement, giving a proportionality between the expectation value of the measured observable and the measurement probabilities of the meter even if $A$ does not commute with $H_0$.

Now, since we wish to perform many measurements in sequence we must find how the system is changed following each measurement step. Defining $\rho_k$ to be the density matrix after $k$ measurements have been performed, we can write
\begin{equation}
    \rho_{k+1} = \begin{cases}
                \frac{\Tr_\mu(\proj_+\alpha_k(\tau))}{P(+1)} & \textrm{with prob. } P(+1)\\
                \frac{\Tr_\mu(\proj_-\alpha_k(\tau))}{P(-1)} & \textrm{with prob. } P(-1)\\
    \end{cases} \label{eq:markov}
\end{equation}
where $\proj_\pm = \frac{1 \pm \sigma^y_a}{2}$ are the projectors onto the eigenstates of the ancilla, and $\Tr_\mu$ is the trace over the ancilla. This is simply an application of Lüder's rule \cite{Svensson13}. As both $P(\pm 1)$ and $\Tr_\mu (\proj_\pm \alpha_k(\tau))$ depend on just $\rho_k$ and not $\rho_{k-1}$ or earlier, this defines a Markov process for $\rho_k$. Consequently, we can apply ideas from discrete-time stochastic analysis to understand the behavior of the system.

For completeness and since it will be used later, the expression for $\Tr_\mu(\proj_\pm \alpha_k(\tau))$ to second order is
\begin{widetext}
\begin{equation}
    \begin{aligned}
        \Tr_\mu(\proj_\pm \alpha_k(\tau)) &= \frac{\rho_k}{2} + \frac{i \tau}{2}[\rho_k,H_0]
        + \frac{\tau^2}{2}(H_0\rho_kH_0 - \frac{1}{2}\{H_0^2,\rho_k\})
        \pm \frac{1}{2}\gamma\tau(\rho_k A + A \rho_k)\\
        &\pm \frac{i\gamma\tau^2}{2} \Bigl( A \rho_k H_0 - H_0 \rho_k A
        + \frac{1}{2}[\rho,AH_0] + \frac{1}{2}[\rho,H_0A]\Bigr)
        + \frac{\gamma^2\tau^2}{2}(A\rho A - \frac{1}{2}\{A^2,\rho\})
    \end{aligned} \label{eq:trmpa}
\end{equation}
\end{widetext}

\subsection{Expectation Value of ``Target'' Observable}

We now introduce another observable $B$ with the requirement that $[B,H_0] = [B,A] = 0$ and investigate how the expectation value of $B$ behaves. Let us examine the quantity $\mathbb{E}(\expval{B}_{k+1}) = P(+1)\Tr(B\rho^+_{k+1}) + P(-1)\Tr(B\rho^-_{k+1})$, where $\rho^\pm_{k+1}$ refers to the two branches of the Markov process defined in (\ref{eq:markov}). This quantity is the expected value of the expectation value of $B$ at step $k+1$, i.e. the ensemble average of $\expval{B}_{k+1}$ if the step $k\rightarrow k+1$ were to be simulated many times. By (\ref{eq:markov}) and (\ref{eq:trmpa}) we obtain
\begin{equation}
    \mathbb{E}(\expval{B}_{k+1}) = \expval{B}_k
\end{equation}
and so the expectation value of $B$ is in on average unaffected by the measurement process.

This generalizes the first property of QND measurement, as choosing $B=A$ simply results in a conserved expectation value of the measured observable. It is then of interest to examine whether the expectation value convergence seen in QND measurement also generalizes in some way.

To do this, we begin by defining $\proj_m$ to be the projector onto the $m^\textrm{th}$ eigensector of $B$, and we define the function
\begin{equation}
    v(\rho) = \sum_{n<m} \sqrt{\Tr(\proj_n \rho)\Tr(\proj_m \rho)} = \sum_{n<m}v_{nm}(\rho).
\end{equation}
The function $v(\rho)$ is 0 if and only if $\rho$ is entirely within one $B$ eigensector, so it serves as a measure of whether collapse has occurred \cite{Rouchon2022}.

Each $v_{nm}$ is a bounded super-martingale, i.e. each has the property that at all times $\mathbb{E}(v_{nm}(\rho_{k+1})) \leq v_{nm}(\rho_k)$. Consequently, by the martingale convergence theorem it is guaranteed that each $v_{nm}$ converges to some stable value after a finite (but unknown) time \cite{stochtext}. We will now support the claim that the $v_{nm}$ are super-martingales, and give a condition for when they all converge to $0$. 
It is evident that $v(\rho)$ is positive semi-definite (showing boundedness), and inspection reveals that $v(\rho) = 0$ if and only if $\rho$ is collapsed into an eigensector of $B$. Thus, if we can show that $\mathbb{E}(v(\rho_{k+1}) - v(\rho_k)) \leq 0$ then $v(\rho)$ is a super-martingale and its convergence will tell us about collapse w.r.t. $B$.

To assess this, we calculate:
\begin{equation}
    \begin{aligned}
    \mathbb{E}(\Delta v_{mn}) &= P(+1)v_{mn}(\rho^+_{k+1}) +\\ &P(-1)v_{mn}(\rho^-_{k+1}) - v_{mn}(\rho_k).
    \end{aligned}
\end{equation}
Substituting in (\ref{eq:trmpa}), we obtain
\begin{widetext}
\begin{equation}
    \begin{aligned}
    &\mathbb{E}(\Delta v_{mn}) = -\frac{\gamma^2\tau^2}{2v_{mn}(\rho)^3}\left(\Tr(\proj_n\rho_k)\Tr(\proj_mA\rho_k) - \Tr(\proj_m\rho_k)\Tr(\proj_nA\rho_k)\right)^2.
    \end{aligned}
\end{equation}
\end{widetext}
Since the term in parentheses is squared and we already know $v_{nm} \geq 0$, it is clear that $\mathbb{E}(\Delta v) \leq 0$ and so $v$ is a super-martingale which converges almost surely. There are then two remaining possibilities concerning the limit of $v$: either $v$ converges to 0 signifying that the state converges into a single eigensector of $B$, or $v$ converges to some other value. The latter case requires that an equilibrium be reached in which the term in parentheses is $0$ at all times. Manipulating this term, we can rephrase this condition as being that the convergence stops at a step labeled $k_0$ when the condition
\begin{equation}
    \Tr\left(\frac{A\proj_n\rho_k\proj_n}{\Tr(\proj_n\rho_k)}\right) = \Tr\left(\frac{A\proj_m\rho_k\proj_m}{\Tr(\proj_m\rho_k)}\right)
    \label{eq:collapsehalting}
\end{equation}
is satisfied for all $n,m$ with $k\geq k_0$. Physically, this condition is that the expectation value of $A$ is at all times independent of any post-selection of an eigenvalue of $B$. To make contact with the QND case we again set $B=A$, giving the condition that $\expval{A}$ must not depend on any post-selection on $A$. This is clearly only met when the state occupies just one eigensector of $A$. So, we recover the QND result of guaranteed convergence to an eigenvalue. We therefore see that this result generalizes the second property of ordinary QND. This means that in our AO-QND protocol, measurement of the auxiliary observable $A$ really does cause the target observable $B$ to behave in a QND manner, thus effectively enabling a measurement-by-proxy of the target observable.

Considering this condition more closely, we can see a direct link between information extraction and collapse. As seen in Eq. \ref{eq:weakprobs}, the information extracted by the weak measurement is information about $\expval{A}$. Rephrasing the collapse-halting condition, we can say that the collapse stops when information about $\expval{A}$ ceases to be correlated with the value of $B$. So, we see that the weak measurement process indirectly extracts information about $B$, and that as long as it does so the system slowly collapses towards a $B$-eigensector. The collapse then stops either when an eigensector is reached and there is no new information about $B$ to be gained, or when eigensectors of $B$ can no longer be distinguished via $\expval{A}$ and so the weak measurement stops extracting new information about $B$. A similar result was obtained in \cite{Andersen22}, showing that an initially mixed state would gradually purify towards a subset of sectors as long as their signatures could be distinguished by the measurement. Here we have shown that this applies for arbitrary initial states, including pure states which are a superposition of contributions from multiple sectors.

It is of further interest to examine what is required for a system to collapse only partially with respect to $B$. By partial collapse, we mean a situation in which the state does not become confined to a single $B$ eigensector, but rather reaches some superposition across multiple sectors which satisfies the collapse-halting condition of Eq. \ref{eq:collapsehalting}. The collapse-halting condition can be rephrased as requiring a state in which $\expval{A}$ is identical in two or more $B$-sectors for all $k>k_0$. That is, supposing we have some state with weight in two distinct $B$ eigensectors and $\ket{b_n},\ket{b_m}$ are the projections of this state in those sectors, $\bra{b_n}A\ket{b_n}_k = \bra{b_m}A\ket{b_m}_k$. This necessarily requires $\expval{A}$ in each of the involved sectors to have the same time-dependence to maintain the balance. There are two sources of time-dependence acting on the system: the unitary evolution under which phases are introduced to energy eigenstates according to their eigenvalue, and the stochastic evolution introduced by the repeated measurements which adjust the weights of different $A$ eigenstates. Assuming that these two sources of time-dependence affect $\expval{A}$ independently over a single measurement cycle since one is deterministic and the other stochastic, both sources of time-dependence must therefore separately match across the involved sectors. To make this more concrete, consider the decomposition of $\ket{b_n}$ into $A$ eigenstates: $\ket{b_n} = \sum_a c_{n,a}\ket{n,a}$, where $\ket{n,a}$ represents a simultaneous eigenstate of $B$ and $A$ in the $n^{\textrm{th}}$ eigensector of $B$ with $A$ eigenvalue $a$. We can further rewrite this as $\ket{b_n} = \sum_a\sum_E c_{n,a}c_{n,a,E}\ket{n,E}$ by expanding the $\ket{n,a}$ into a sum over simultaneous $B$ and energy eigenstates in the same sector. We can now argue that over some small timestep the Hamiltonian time-dependence acts only on $c_{n,a,E}$ by introducing a phase, while the time-dependence from weak measurement acts only on the $c_{n,a}$ by nudging these weights slightly in favor of some $a$ which is determined by the random measurement outcome. Thus, if we wish to find a non-collapsing state with weight in two sectors $n,m$, the decomposition in both sectors should be identical in the sense that $c_{n,a} = c_{m,a}$ and $c_{n,a,E} = c_{m,a,E}$. We have thus far not found any nontrivial examples of such a condition being met.

\section{Application to Heisenberg Chain}
\begin{figure*}[htb]
    \subfloat{
        \begin{tikzpicture}[node distance=0cm]
        \node (img) {\includegraphics[width=0.35\textwidth]{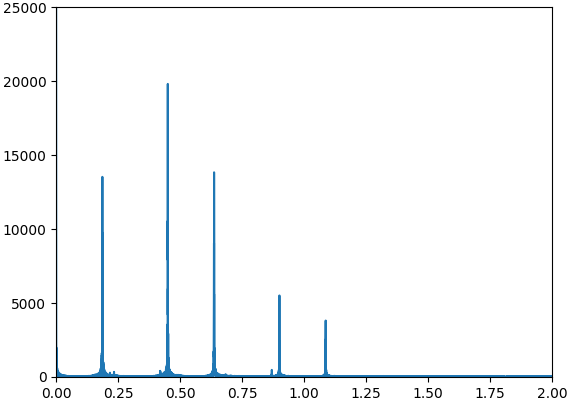}};
        \node[above left=of img] {(a)};
        \node[left=of img, xshift=-0.1cm,rotate=90,anchor=center] {\footnotesize Amplitude};
        \node[below=of img]{\footnotesize Frequency ($J/2\pi$)};
        \node[left=of img,xshift=1.8cm,yshift=0.4cm] {\footnotesize 2$\leftrightarrow$1a};
        \node[left=of img,xshift=2.51cm,yshift=1.72cm] {\footnotesize 1a$\leftrightarrow$1b};
        \node[left=of img,xshift=2.51cm,yshift=1.45cm] {\footnotesize 1b$\leftrightarrow$1c};
        \node[left=of img,xshift=3.05cm,yshift=0.5cm] {\footnotesize 2$\leftrightarrow$1b};
        \node[left=of img,xshift=3.73cm,yshift=-0.8cm] {\footnotesize 1a$\leftrightarrow$1c};
        \node[left=of img,xshift=4.25cm,yshift=-1.15cm] {\footnotesize 2$\leftrightarrow$1c};
        \end{tikzpicture}
    }%
    \subfloat{
        \begin{tikzpicture}[node distance=0cm]
        \node (img) {\includegraphics[width=0.35\textwidth]{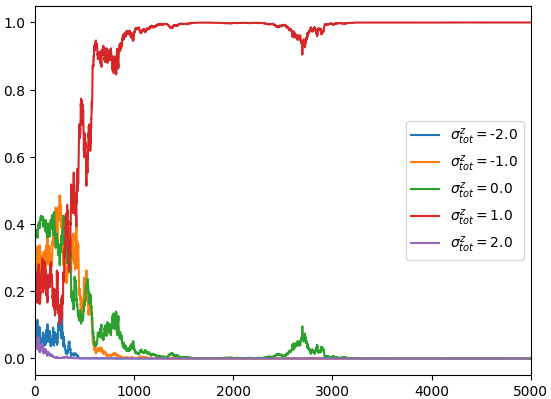}};
        \node[above left=of img] {(b)};
        \node[left=of img,xshift=-0.2cm,rotate=90,anchor=center]{\footnotesize Overlap with $\sigma^z_{tot}$ Sector};
        \node[below= of img] {\footnotesize Timestep};
        \end{tikzpicture}
    }
    
    \subfloat{
        \begin{tikzpicture}[node distance=0cm]
        \node (img) {\includegraphics[width=0.4\textwidth]{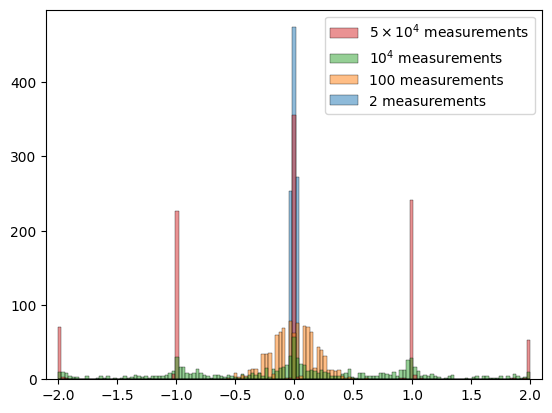}};
        \node[below= of img] {\footnotesize $\expval{\sum_i \sigma^z_i}$};
        \node[above left=of img, yshift=-0.5cm] {(c)};
        \node[left=of img,xshift=-0.2cm,rotate=90,anchor=center]{\footnotesize Count};
        \end{tikzpicture}
    }
    \caption{Behavior of the 4-site Heisenberg chain subjected to repeated weak measurement of $\sigma^z_0$. a) Fourier Transform of $\expval{\sigma^z_0}(t)$ showing sharp peaks corresponding to certain energy gaps in the spectrum of the chain. Labels on each peak show the contributing components, as identified in Table 1. b) Overlap of the chain's state with sectors of total z-component shows a collapse into one sector following stochastic evolution. c) Distribution of expectation values of total z-component for a sample of 1,000 chains shown after certain numbers of measurements. It is clear that each individual run collapses into an eigensector of total z-component, while the ensemble average expectation value remains 0, demonstrating QND-like behavior.}
    \label{fig:hbergsectors}
\end{figure*}

Now we present a concrete example where the convergence of a secondary observable occurs. The system we examine here is a Heisenberg chain with the Hamiltonian:
\begin{equation}
    H = -J \sum_{ij}\vec{\sigma}_i\cdot \vec{\sigma}_j
\end{equation}
We choose to label the spins along the chain from $0$ to $N$ so that e.g. $\vec{\sigma}_0$ refers to the first spin on the chain, the ancilla. The operator we weakly measure is $\sigma^z_0$, the z-component of the first spin. To make the connection clear, this $\sigma^z_0$ operator corresponds to $A$ in the previous section and clearly does not commute with the Heisenberg Hamiltonian. The operator corresponding to $B$ will be the total z-component of the chain, which we refer to here as $m_s$. It is easy to verify that the required condition $[A,B] = [B,H] =0$ is met by this set of operators. The experimentally accessible information from this protocol will be the weak measurement record of $+1$ and $-1$ outcomes of the ancilla, but here we also explicitly track the system's state over time in order to gain more insight into how it behaves.

Figure \ref{fig:hbergsectors} shows results for a 4-spin chain, showing that after an initial period of stochastic evolution, the state of the chain eventually collapses into a single $m_s$ sector. This behavior mimics QND collapse such as that seen in figure \ref{fig:qnd} and supports the theory we developed in the previous section.

Of additional interest is the Fourier transform of $\expval{\sigma^z_0}$, which reveals oscillatory behavior in the z-component of the measured spin with several sharp peaks in frequency space. Examination of these peaks reveals that their spacing corresponds to certain spacings in the energy spectrum of the Heisenberg chain. More specifically, the oscillation frequencies correspond to energy spacings between eigenstates with the same total z-component (hence inhabiting the same $B$-sector) but with different energies and potentially different total spin, with allowed peaks following the selection rule $\Delta S = 0,1$. In particular, the frequencies seen in Fig. 2a represent such spacings between eigenstates in the $m_s=1$ sector. That is, pairs of eigenstates with $m_s=1$ for both states and a difference of 0 or 1 in total spin contribute to the oscillation of the measured end spin. The fact that all possible peaks allowed under this rule are present indicates that following collapse into a total z-component sector, the effect of the measurement back-action causes the system state to explore all available states within the sector. It is yet to be determined whether this complete exploration of the sector is guaranteed for other systems, but there must be at least a partial exploration of the sector since by construction the measurement doesn't commute with the Hamiltonian and so must perturb the system's state to some extent. The finding of frequency peaks which obey this particular selection rule is specific to the Heisenberg chain but supports the finding of collapse into a single $m_s$ sector, as otherwise additional peaks would be present. These oscillations would also be seen in the experimentally accessible measurement record, though suppressed by noise. As such, our findings can be experimentally tested as long as a sufficient quantity of data is collected to overcome the statistical penalty of weak measurement. It is not clear at this time whether our protocol guarantees a similar experimental signature for arbitrary systems.

\begin{table}[h]
    \centering
    \begin{tabular}{|c|c c c c c c|}
        \hline
        \textbf{Total Spin Sector} &  2 & 1a & 0a & 1b & 1c & 0b\\
        \hline
        \textbf{Total Spin} & 2 & 1 & 0 & 1 & 1 & 0\\
        \hline
        \textbf{Energy} & -3 & -1.83 & -0.46 & 1 & 3.83 & 6.46\\
        \hline
    \end{tabular}
    \caption{Total spin sectors of the 4-site Heisenberg chain and their energies (J=1).}
    \label{tab:my_label}
\end{table}

\FloatBarrier

\section{Conclusion and Outlook}
Our key result is that repeated weak measurements of a non-conserved observable can extract information about other observables and lead to a partial collapse of the system's density matrix, giving rise to a new potential class of measurement protocols we term Auxiliary Observable QND. On one hand, this is intriguing because it sheds more light on the measurement problem and the nature of wavefunction collapse, reinforcing the idea that a quantum system tends towards collapse whenever information is extracted from it, even if such information is gained indirectly. On the other hand, this result also presents a new approach to measurement that may have some utility in experiment. For example, if an observable of interest is difficult to measure directly it may be possible to find a different observable which is easier to measure yet still indirectly extracts the desired information. This is highlighted in the Heisenberg chain example, where measurement of a local observable was able to reveal information about a total system observable. In particular, such a protocol may find a useful application in Rydberg array experiments where coupling ancilla qubits to one or a few sites is a very natural operation, so this type of ancilla-based indirect measurement may be easier to implement than a direct measurement.

\FloatBarrier

\begin{acknowledgments}
This work was supported by National Science Foundation Division of Mathematical Sciences Grant No. 2138905.
\end{acknowledgments}

\bibliography{refs}

\end{document}